\documentstyle[aps,preprint,epsf]{revtex}
\newcommand{\be}{\begin{equation}}
\newcommand{\ee}{\end{equation}}
\begin{document}
\preprint{SNUTP}
\draft
\title{Island Density in Homoepitaxial Growth:\\ 
Improved Monte Carlo Results \\} 
\author {H.~Jeong$^1$, B.~Kahng$^2$, and D.~E.~Wolf$^3$ \\}
\address {$^1$ Department of Physics and Center for Theoretical 
Physics, \\ 
Seoul National University, Seoul 151-742, Korea \\ 
$^2$ Department of Physics and Center for Advanced Materials 
and Devices, \\ 
Kon-Kuk University, Seoul 143-701, Korea \\
$^3$ FB 10, Gerhard-Mercator Universit\"at, D-47048 Duisburg, Germany \\} 

\maketitle
\thispagestyle{empty}

\begin{abstract}
We reexamine the density of two dimensional islands in the submonolayer  
regime of a homoepitaxially growing surface using 
the coarse grained Monte Carlo simulation with random 
sequential updating rather than parallel updating.     
It turns out that the power law dependence of 
the density of islands on the deposition rate 
%obtained by random sequential updating 
agrees much better with the theoretical prediction than previous 
data obtained by other methods if random sequential instead of 
parallel updating is used. 
\end{abstract} 

          
%\begin{multicols}{2}
\narrowtext 

Recently problems of surface growth by molecular beam epitaxy (MBE) have 
attracted a lot of interest \cite{family,markov,lagally,godreche}. From the point of view of statistical
physics it is intriguing, how various stochastic processes such as
shot noise, diffusion, and nucleation, give rise to scaling structures involving many atoms.
These cooperative phenomena depend crucially on the ratio between the 
diffusion constant $D$ and the deposition rate $F$ 
(number of atoms landing on the surface
per unit area and unit time).
If adatoms can diffuse to  stable positions
such as kink sites or step edges, before other adatoms get deposited,
and if interlayer diffusion is not inhibited by Ehrlich-Schwoebel 
barriers \cite{ehrlich,schwoebel},
smooth surfaces with a minimum of defects can be grown. 
High symmetry surfaces
exhibit layer-by-layer growth called also
Frank-van-der-Merve mode \cite{schommers}. However, when the deposition 
rate is high 
compared to the hopping rate of adatoms,the
surface gets rough quickly.

In the present note we reconsider the structures formed
when less than a monolayer is deposited. It has long been known 
that the maximum density of islands, before coalescence starts, 
has a power law dependence $\rho \propto (D/F)^{-2\gamma}$
\cite{zinsmeister,venables,vpw,vptw}. 
Using rate equations, the exponent $\gamma$ 
can be predicted.
If only adatoms can move and desorption can be neglected, 
one obtains \cite{vpw,vptw}
\be
\gamma = {{ i^*} \over  { 2 i^* + d + d_f } },
\ee 
where $s^*=i^*+1$ is the size of the smallest stable island, 
which means that an island with more than $i^*$ atoms is more 
likely to grow than to shrink. $d$ is the surface dimension
and $d_f$ the fractal dimension of the islands \cite{pimpinelli}.

This prediction has been supported by many computer simulations for various 
models \cite{evans,tang,pvw,smilauer,schroeder}. In particular for 
large $i^*$ the exponents were found to be slightly (up to 10\%) smaller than
the theoretical prediction. The question has been raised, whether this 
must be so, because islands smaller than $s^*$ may also be stable with a 
rather high probability, so that the exponent might be determined by an
effective $i^*_{\rm eff}<i^*$. Here we present new data which indicate that
with a slight modification of the simulation model used in \cite{schroeder}
the $\gamma$-values turn out to be much closer to the theoretical 
prediction than before.

% describe the model

Numerical simulations of MBE require large
computation times, 
because the deposition of one monolayer takes very long on the time scale of  
the adatom diffusion.
Moreover, it is easy to see that the system size required in order to 
avoid strong finite size effects, must be larger than 
$\max(\ell,\ell_0)$, where $\ell \sim (D/F)^{\gamma}$
is the characteristic distance between neighboring islands and
$\ell_0 = (D/F)^{1/(2+d)}$ is the only combination of $F$ and $D$ with 
length dimension. If the system is smaller than $\ell_0$,
it can accommodate at most one island, so that the island density cannot 
be obtained \cite{schroeder}. Recently it was shown \cite{kallabis}
that there is another characteristic length in the system, called the
layer-coherence length $\tilde\ell=\ell^{4/(4-d)}$, which is significantly 
larger than $\max(\ell,\ell_0)$. Systems smaller than
$\tilde\ell$  exhibit perfect layer-by-layer growth, whereas larger surfaces
show kinetic roughening. 
In our computer simulations we  chose the system size larger than $\tilde\ell$
to prevent these finite size effects. Therefore it is essential to have
an efficient simulation model.

Recently a very efficient coarse-grained Monte Carlo (CGMC) 
simulation method has been introduced \cite{wolf}.
The basic idea of the CGMC simulation method is that 
in order to measure the island density one needs not be able
to follow the adatom diffusion on the atomic scale. Instead,
one monitors  the position of adatoms only on a much coarser
scale, $\Delta x$, which must be smaller than the average
distance between islands.
Then, the computation time is 
reduced by about a factor $(\Delta x)^{-2}$.

We consider $L \times L$ cells on a two dimensional substrate, 
each of which is composed of $\Delta x \times \Delta x$ sites.
For $i^*=1, 2$ we used $L=400$ and $\Delta x=2$, whereas for
$i^*=3, 4$ the values were 200 and 4. 
Deposition of an atom into any of the cells happens with frequency 
$\nu_F=F(L\Delta x)^2$. With frequency $\nu_D= D(\Delta x)^{-2}$ all 
adatoms in the system are allowed to move to any of their neighboring cells 
or to stay where they are, with equal probability. If there are no adatoms
in the system, $\nu_D=0$. In each Monte-Carlo step either an atom is 
deposited with
probability $\nu_F/(\nu_F+\nu_D)$, or all adatoms are allowed to move.
The motion of adatoms within the cells is ignored 
in the coarse grained  picture. Instead, the state of each cell is
characterized by three variables: First, $h(x)$ is the number of
completed monolayers which corresponds to the height in cell $x$. 
Second, $0\leq m(x)< (\Delta x)^2$ is the number of
atoms on top of the $h(x)$ monolayers. Finally 
we assign a step indicator ``flag" to each cell, 
which is either 0 or 1, indicating whether the $m$ atoms belonging to 
the cell are mobile or immobile, respectively.
They are immobile, if the cell contains an island edge, because for simplicity 
we assume that then the adatoms are irreversibly bound there, before they 
move out of the cell again.
An island edge can be created within a cell, if $i^*+1$ atoms 
are gathered there, so that a stable island nucleates. 
It may also happen that an island
fills a whole cell and starts invading the neighboring cells, whose step 
indicator is  then set equal to 1 without nucleation.

In a deposition step, a cell is chosen at random, and its number $m$
of atoms incremented by one. If $m$ then exceeds $i^*$, 
a nucleation event is recorded, and the flag is set equal 1. 
The diffusion process can be
implemented in two different ways. In previous work all adatoms
were simultaneously allowed to move into neighboring cells (parallel 
updating of the list of mobile atoms). Afterwards
it was checked, which of the cells contained more than $i^*$ atoms, and
the corresponding  nucleation events were recorded.

Here, we implement the diffusion process in a different way.
As before the mobile atoms are elements of a list. 
However, now starting with the first element the atoms are allowed 
to diffuse to a
neighboring cell sequentially: After each move it is checked whether
it results in a nucleation event, i.e. whether $m$ exceeds $i^*$.
In this case all $m$ atoms form a stable island and are deleted from the 
list of mobile atoms. The holes in the list are filled with the last
elements, which leads to a randomization of the list. 
Then the next atom on the list is
allowed to move, it is checked, whether this triggers a nucleation,
and the list of mobile atoms is updated accordingly. In this way
we continue until we went though the whole list once. 

\begin{figure}
\centerline{\epsfxsize=8cm \epsfbox{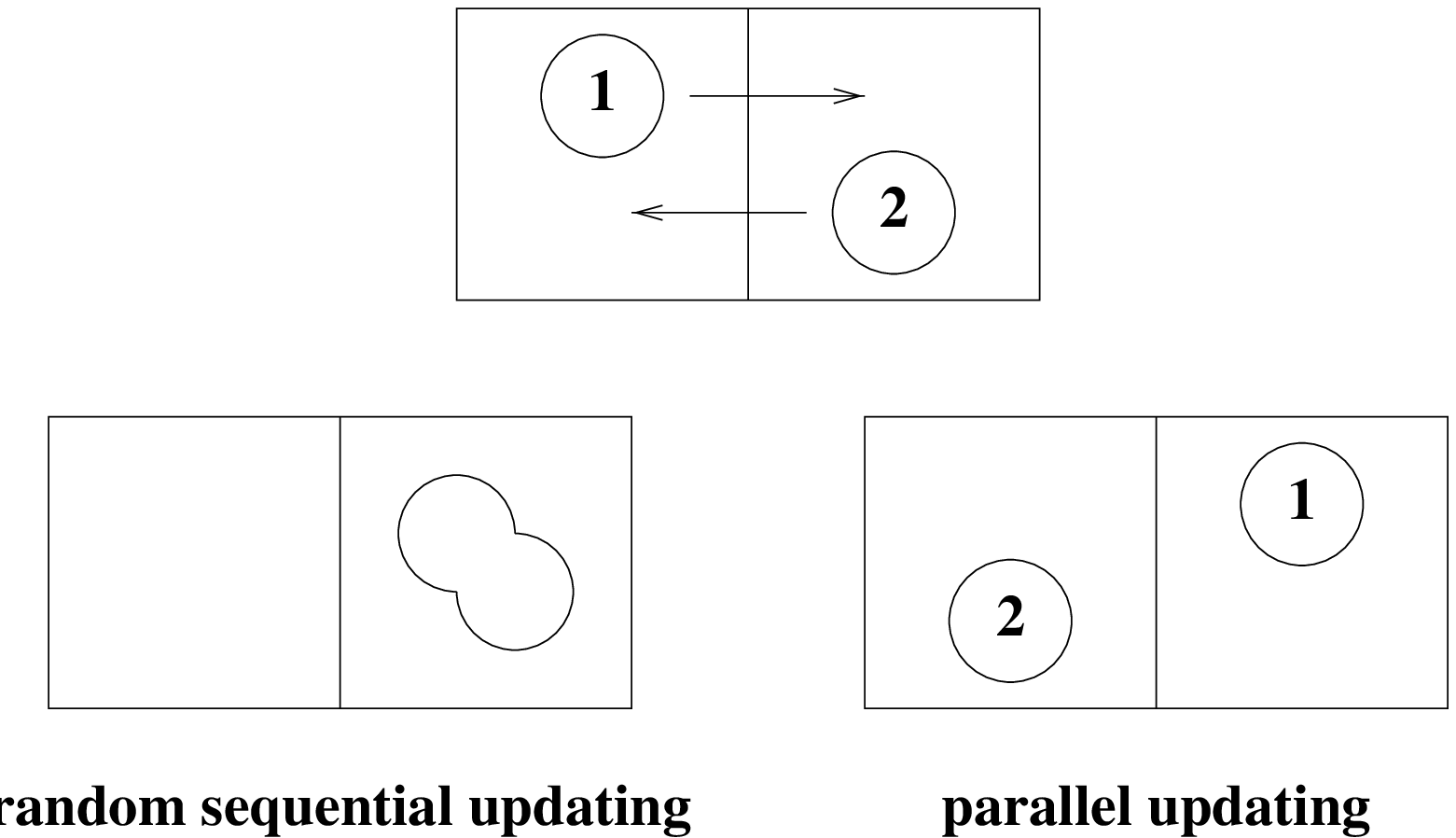}}
\vspace{.5cm}
\caption{ Difference between the random sequential 
and the parallel updating rule.
In the random sequential updating, particle 1 moves first and meets particle 2,
and nucleation occurs ($i^*=1$). But in the parallel updating, particles can 
collide but cross each other without nucleation. }
\label{fig1}
\end{figure}

Fig.\ref{fig1} illustrates the difference between the parallel and the
random sequential way of updating the list of mobile atoms for the case 
$i^*=1$:
For parallel updating it may happen that two atoms just exchange their
positions. Instead, in reality they are so close to each other that 
they could easily meet to form an island. 
The random sequential update takes this into account.

% don't forget to explain how we measure the island size.

In our simulations we recorded the number of islands at a 
coverage 0.12.
The two different ways of updating in the diffusion process give
different values for the exponent $\gamma$. 
It turns out that the random sequential update produces values, which are 
closer to the theoretical ones 
than those obtained by the parallel updating 
\cite{schroeder} and other methods \cite{amar}, especially for 
larger stable island sizes.
The numerical values we obtained are compared  with other values 
in the following table. 
 
%--------------------------------------------------------------------------
%   
%{{ i^* }} *  [1]         [2]         [3]         [4]       [5]      [6] 
%--------------------------------------------------------------------------
%                                                                          
%    1        0.333       0.35         0.344       0.336    0.33    0.336  
%                                                                          
%    2        0.500       0.52         0.48        0.512    0.5     0.523  
%                                                                          
%    3        0.600       0.62         0.55        0.609    0.58    0.592  
%                                                                          
%    4        0.667       0.68         0.58        0.658            0.650  
%                                                                          
%------------------------------------------------------------------------- 
%
In the table,  the values in the column (1) are from Eq.~(1)  
with $d_f = 2$. The values in the column (2) are from Eq.~(1) with 
$d_f=1.7$. The values in the column (3) are the ones from Ref.~8 
using the parallel updating. 
The values in the column (4) are the one obtained by  
Family and  Amar \cite{amar} using the kinetic Monte Carlo method 
\cite{vvedensky}.   
The values in the column (5) are the ones we  
obtained using the random sequential updating and a linear fit to
the double-logarithmic plot of the data (see Fig. 2) excluding the 
ones for $D/F \leq 10^6$. Column (6) contains a fit of the
same data points with the formula
\be
\rho = {\rm const}(D/F)^{-2\gamma}(\log D/F)^{-1/3},
\ee
which shows that the logarithmic correction worked out by Tang \cite{tang}
does not change our conclusion that the random sequential update 
gives better agreement with the theoretical values for the exponent. \\

In summary, we have investigated the island density in the 
submonolayer regime of homoepitaxial growth. The island
density on a two dimensional substrate decreases 
with increasing the ratio $D/F$ 
as $\rho \sim (D/F)^{-2 \gamma}$.  
The exponent $\gamma$ depends on the critical size  
$i^*$.  Performing the coarse grained  Monte 
Carlo simulation with random sequential updating in the diffusion step, 
we have obtained numerical values of the exponent 
$\gamma(i^*)$ for  $i^*$=1, 2, 3, and 4 which are in better agreement 
with the theoretical values than the ones obtained previously 
by other methods. \\

One of authors (H.J.) would like to thank the DaeWoo Foundation for
financial support and SNU computer center for the computing time on SP2.
 B.K. is supported in part by the 
KOSEF (971-0207-025-2), in part by the BSRI program
of Ministry of Education (BSRI-96-2409), and 
in part by the KOSEF through the SRC program of SNU-CTP.
D.W. thanks the Deutsche 
Forschungsgemeinschaft for support within SFB 166.\\

\begin{table}
\begin{center}
\begin{tabular}{ccccccccc}
 & $i^*$ & (1) & (2) & (3) & (4) & (5) & (6) &\\
\hline
 &1 & 0.333 & 0.35 & 0.344 & 0.33 & 0.336 & 0.336 &\\
 &2 & 0.500 & 0.52 & 0.48 & 0.5 & 0.507 & 0.523 &\\
 &3 & 0.600 & 0.62 & 0.55 & 0.58 & 0.609 & 0.592 &\\
 &4 & 0.667 & 0.68 & 0.58 &  & 0.658 & 0.650 &\\
\end{tabular}
\end{center}
\caption{The  comparison  of  the numerical  values  of   
{{ $2\gamma$ }} with  the 
theoretical values and  the ones obtained by other simulation 
methods.
The values in the column (1) are from the theoretical formular
Eq.~(1) with $d_f=2$. The values in the column (2) are from Eq.(1)
with $d_f=1.7$. The values in the column (3) are quoted from
Ref.8 using the parallel updating. The values in the column (4)
are quoted from Ref.20 using the kinetic Monte Carlo
method. The values in the column (5) are the ones we obtained
using the random sequential updating.
The values in the column (6) are the ones fitted by the formular
(2) with the logarithmic correction.}
\end{table}

\begin{figure}
\centerline{\epsfxsize=8cm \epsfbox{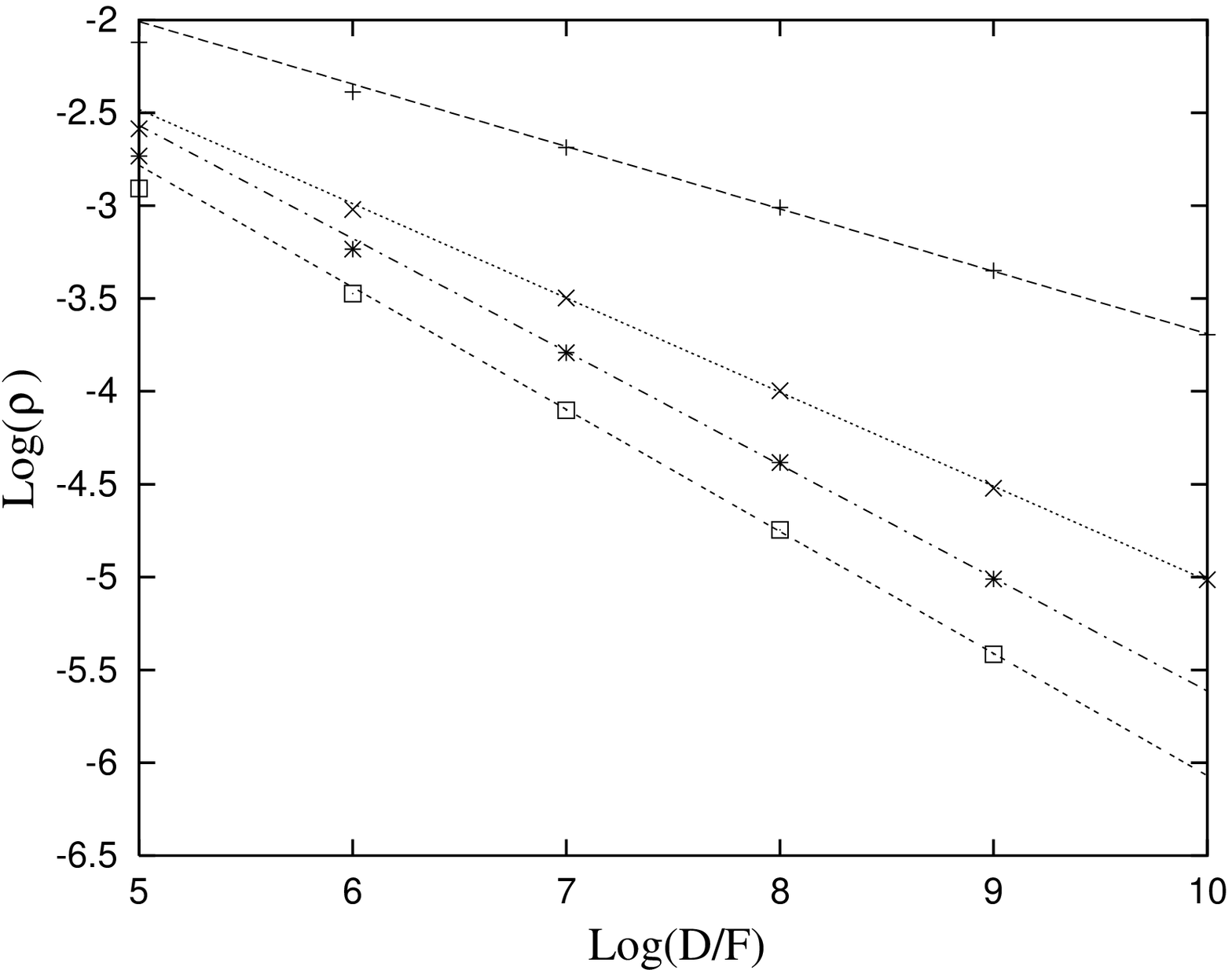}}
\vspace{.5cm}
\caption{The density of islands versus the ratio   
$D/F$ in log-log scale for the cases 
of $i^*=1, 2, 3,$ and 4 from the top.}
\label{fig2}
\end{figure}

%\end{multicols}


\begin{references} 

\bibitem{family} F. Family and T. Vicsek, {\it Dynamics of Fractal 
Surfaces} (World-Scientific, Singapore, 1991). 
\bibitem{markov} I.V. Markov, {\it Crystal Growth for Beginners} 
(World-Scientific, Singapore, 1995).
\bibitem{lagally} M.G. Lagally, {\it Kinetic of Ordering and Growth at Surfaces}
(Prenum Press, New York, 1990).
\bibitem{godreche} C. Godreche, {\it Solids far from Equilibrium} 
(Cambridge University Press, New York, 1992). 
\bibitem{ehrlich} G. Ehrlich and F. Hudda, J. Chem. Phys. {\bf 44}, 1039 (1966).
\bibitem{schwoebel} R.L. Schwoebel, J. Appl. Phys. {\bf 40}, 614 (1969).
\bibitem{schommers} W. Schommers and P. von Blanckenhagen, {\it Structure 
and Dynamics of Surfaces II} (Springer-Verlag, Berlin, 1987).
\bibitem{zinsmeister} G. Zinnsmeister, Thin Solid Films {\bf 2}, 497 (1968); 
                                      {\em ibid} {\bf 7}, 51 (1971).
\bibitem{venables} J.A. Venables, G.D. Spiller, M. Hanb\"ucken, Rep. Prog. Phys. {\bf 47}, 399 (1984).
\bibitem{vpw} J. Villain, A. Pimpinelli, D.E. Wolf, Comments Cond. Mat. Phys. {\bf 16}, 1 (1992).
\bibitem{vptw} J. Villain, A. Pimpinelli, L.-H. Tang, D.E. Wolf, J. Phys. I (France) {\bf 2}, 2107 (1992).
\bibitem{pimpinelli} A. Pimpinelli, J. Villain, and D.E. Wolf, 
J. Phys. I (France) {\bf 3}, 447 (1993). 
\bibitem{evans} J.W. Evans, M.C. Bartelt, J. Vac. Sci. Technol. A {\bf 12}, 1800 (1994).
\bibitem{tang} L.H. Tang, J. Physique I {\bf 3}, 935 (1993).
\bibitem{pvw} A. Pimpinelli, J. Villain, D.E. Wolf, J. Phys. I (France) {\bf 3}, 447 (1993); Phys. Rev. Lett.
 {\bf 69}, 985 (1992).
\bibitem{smilauer} C. Ratsch, A. Zangwill, P. Smilauer, D.D. Vvedensky,
               Phys. Rev. Lett. {\bf 72}, 3194 (1994).
\bibitem{schroeder} M. Schroeder and D.E. Wolf, Phys. Rev. Lett. {\bf 74}, 2062 (1995).
\bibitem{kallabis}  H. Kallabis, L. Brendel, J. Krug, D.E. Wolf,
            Intl. J. Mod. Phys. B, (to be published).
\bibitem{wolf} D.E.Wolf,  in {\it Scale invariance, Interface, and 
Non-Equilibrium Dynamics}, edited by M. Droz, A.J. McKane, 
J. Vannimenus and D.E. Wolf, Nato-ASI Series (Plenum, New York, 1994).
\bibitem{amar} J.G. Amar and F. Family, Phys. Rev. Lett. {\bf 74}, 2066 (1995).
\bibitem{vvedensky} C. Ratsch, A. Zangwill, P. Smilauer, D.D. Vvedensky, Phys. Rev. Lett. {\bf 72} 3194 (1994).
\end{references}
\end{document}